\documentclass[runningheads,a4paper,10pt]{llncs}

\usepackage{graphicx}
\usepackage{caption}
\usepackage[hyphens]{url}
\usepackage{xspace}
\usepackage{authblk}
\usepackage{multirow}
\usepackage{longtable}
\usepackage[normalem]{ulem}
\usepackage{listings}
\usepackage{xcolor}
\usepackage[utf8]{inputenc}
\definecolor{mygreen}{rgb}{0,0.6,0}
\definecolor{mygray}{rgb}{0.5,0.5,0.5}
\definecolor{mymauve}{rgb}{0.58,0,0.82}
\definecolor{base03}{HTML}{002b36}
\definecolor{base02}{HTML}{073642}
\definecolor{base01}{HTML}{586e75}
\definecolor{base00}{HTML}{657b83}
\definecolor{base0}{HTML}{839496}
\definecolor{base1}{HTML}{93a1a1}
\definecolor{base2}{HTML}{eee8d5}
\definecolor{base3}{HTML}{fdf6e3}
\definecolor{yellow}{HTML}{b58900}
\definecolor{orange}{HTML}{cb4b16}
\definecolor{red}{HTML}{dc322f}
\definecolor{magenta}{HTML}{d33682}
\definecolor{violet}{HTML}{6c71c4}
\definecolor{blue}{HTML}{268bd2}
\definecolor{cyan}{HTML}{2aa198}
\definecolor{green}{HTML}{859900}
\newcommand{\codefont}{\fontfamily{pcr}\selectfont}

\lstdefinelanguage{JavaScript}{
  keywords={function, var, const, class, default, typeof, void, export, extends, import, super},
  keywordstyle=\color{blue}\bfseries,
  ndkeywords={break, case, try, catch, continue, debugger, delete, do, else, export, throw, while,
  with, yield, for, finally, if,  in, instanceof, new, return, switch, null},
  ndkeywordstyle=\color{yellow}\bfseries,
  identifierstyle=\color{base02},
  sensitive=false,
  comment=[l]{//},
  morecomment=[s]{/*}{*/},
  commentstyle=\color{base1},
  stringstyle=\color{cyan}\bfseries,
  morestring=[b]',
  morestring=[b]"
}

\lstdefinelanguage{HTML}{
  keywords={a, abbr, address, area, article, aside, audio, b, base, bdi,
  bdo, blockquote, body, br, button, canvas, caption, cite, code, col, colgroup, datalist, dd, del,
  details, dfn, dialog, div, dl, dt, em, embed, fieldset, figcaption, figure, footer, form, h1, h2, h3, h4, h5, h6, head, header, hr, html, i, iframe, img, input, ins, input, ins, kbd, keygen,
  label, legend, li, link, main, map, mark, menu, menuitem, meta, meter, nav, noscript, object, ol, optgroup, option, output, p, param, picture, pre, progress, q, rp, rt, ruby, s, samp, script, section, select, small, source, span, strong, style, sub, summary, sup, table, tbody, td,
  textarea, tffot, th, thead, time, title, tr, track, u, ul, video, wbr},
  keywordstyle=\color{blue}\bfseries,
  ndkeywords={break, case, catch, class, const, continue, debugger, default, delete, do, else, export, extends, finally, for, function, if, import, in, instanceof, new, return, super, switch, this, throw, try, typeof, var, void, while, with, yield, ||, &&},
  ndkeywordstyle=\color{yellow}\bfseries,
  identifierstyle=\color{base02},
  sensitive=false,
  comment=[l]{//},
  morecomment=[s]{<!--}{-->},
  morecomment=[n]{/*}{*/},
  commentstyle=\color{base1},
  stringstyle=\color{cyan}\bfseries,
  morestring=[b]',
  morestring=[b]"
}

\newcommand{\accesstoken}{\textit{access\_token}\xspace}

\newcommand{\idtoken}{\textit{id\_token}\xspace}

\newcommand{\redirecturi}{\textit{redirect\_uri}\xspace}
\begin{document}
\title{OAuthGuard: Protecting User Security and Privacy with OAuth 2.0 and OpenID Connect}
\author{Wanpeng Li\inst{1} \and Chris J Mitchell\inst{2} \and Thomas Chen\inst{3}}
\titlerunning{Protecting User Security and Privacy with OAuth 2.0 and OpenID Connect}
\institute{
School of Computing, Mathematics and Digital Technology\\
Manchester Metropolitan University, UK \\
\email{W.Li@mmu.ac.uk}
  \and
  Information Security Group\\
  Royal Holloway,
  University of London, UK \\ \email{me@chrismitchell.net}
  \and
  Department of Electrical \& Electronic Engineering\\
  City, University of London, UK \\ \email {Tom.Chen.1@city.ac.uk}
}

\maketitle

\begin{abstract}
Millions of users routinely use Google to log in to websites supporting OAuth
2.0 or OpenID Connect; the security of OAuth 2.0 and OpenID Connect is
therefore of critical importance. As revealed in previous studies, in practice
RPs often implement OAuth 2.0 incorrectly, and so many real-world OAuth 2.0 and
OpenID Connect systems are vulnerable to attack. However, users of such flawed
systems are typically unaware of these issues, and so are at risk of attacks
which could result in unauthorised access to the victim user's account at an
RP\@. In order to address this threat, we have developed \emph{OAuthGuard}, an
OAuth 2.0 and OpenID Connect vulnerability scanner and protector, that works
with RPs using Google OAuth 2.0 and OpenID Connect services. It protects user
security and privacy even when RPs do not implement OAuth 2.0 or OpenID Connect
correctly. We used OAuthGuard to survey the 1000 top-ranked websites supporting
Google sign-in for the possible presence of five OAuth 2.0 or OpenID Connect
security and privacy vulnerabilities, of which one has not previously been
described in the literature. Of the 137 sites in our study that employ Google
Sign-in, 69 were found to suffer from at least one serious vulnerability.
OAuthGuard was able to protect user security and privacy for 56 of these 69
RPs, and for the other 13 was able to warn users that they were using an
insecure implementation.


\end{abstract}
\section{Introduction} 
\label{sec:introduction}

Since the OAuth 2.0 authorisation framework was published at the end of 2012
\cite{oauth2}, it has been adopted by a many websites worldwide as a means of
providing single sign-on (SSO) services. By using OAuth 2.0, websites can
reduce the burden of password management for their users, as well as saving
users the inconvenience of re-entering attributes that are instead stored by
identity providers and provided to relying parties as required. There is a
correspondingly rich infrastructure of identity providers (IdPs) providing
identity services using OAuth 2.0. Indeed, some relying parties (RPs), such as
the website
USATODAY\footnote{\url{https://login.usatoday.com/USAT-GUP/authenticate/?}},
support as many as six different IdPs.


The security of OAuth 2.0 and OpenID Connect is therefore of critical
importance, and it has been widely examined both in theory and in practice.
Previous studies show that, in practice, RPs do not always implement OAuth 2.0
correctly; as a result, many real-world OAuth 2.0 and OpenID Connect systems
are vulnerable to attack. Researchers have developed a range of mitigations for
RP developers, designed to help secure OAuth 2.0 and OpenID connect systems.
However, none of this prior art is aimed at protecting users who are
(unwittingly) employing an insecure OAuth 2.0 or OpenID Connect implementation.

To close this gap, we have developed \emph{OAuthGuard}, an
OAuth 2.0 and OpenID Connect vulnerability scanner and
protector, for use with RPs using Google OAuth 2.0 and OpenID
Connect services. While we have focussed only on Google sign-in
in the work described here, we believe that the same approach
can be used to protect user security and privacy when working
with other identity providers.

\label{sub:our_contribution}

The main contributions of this paper are as follows.
\begin{enumerate}
  \item \textbf{A new vulnerability}.  We identify a new
      privacy vulnerability which is present in a number of
      real-world websites.
  \item \textbf{OAuthGuard}. We describe the design and implementation of
      OAuthGuard (see Section \ref{sec:oauthguard}), which provides
      real-time protection for users against vulnerabilities arising from
      poor implementations of OAuth 2.0 and OpenID Connect by web sites
      (RPs) using the Google SSO service. Despite the ubiquity of these
      implementation vulnerabilities, this is the first practical help that
      has been offered to end users. We also outline how we addressed the
      challenges we faced in making the system work, including the
      trade-offs we made to ensure that OAuthGuard is compatible with all
      the RPs in our study.

  \item \textbf{A large-scale study}. We ran OAuthGuard on the top 1,000
      websites from
      \url{majestic.com}\footnote{\url{https://majestic.com/reports/majestic-million}}
      (Section \ref{sec:case_study}). Key results from the study include
      finding at least one vulnerability in 69 of the 137 RPs that use
      Google Sign-in (Section \ref{sub:study_result}). We further manually
      analysed the 109 RPs in the top 1,000 for which OAuthGuard did not
      detect a CSRF attack threat, and found that 25 of them are
      nevertheless vulnerable to a CSRF attack. Of the 69 RPs it found to
      be vulnerable, OAuthGuard is able to protect users against CSRF
      attacks for 48 of the 53 RPs (91\%) which are vulnerable to such an
      attack; OAuthGuard was also able to upgrade the protocol from HTTP to
      HTTPS for 8 of the 13 RPs (62\%) that erroneously use HTTP to
      transfer their OAuth 2.0 response. OAuthGuard identified nine RPs
      that leak user tokens to third party websites, either unintentionally
      or intentionally, and in total blocked 75 http requests leaking user
      tokens for these nine RPs.  Finally, OAuthGuard generated a warning
      to users for 13 RPs that are vulnerable to an impersonation attack.
\end{enumerate}


The remainder of this paper is structured as follows.  Section
\ref{sec:background} provides background on OAuth 2.0 and OpenID Connect.
Section \ref{sec:related_work} describes related work. In Section
\ref{sec:vulnerabilities}, we describe the five vulnerabilities that OAuthGuard
can detect and mitigate, one of which was not previously known. Section
\ref{sec:oauthguard} specifies the infrastructure of OAuthGuard. In Section
\ref{sec:case_study}, we describe a case study on the Google sign-in security
of 137 RPs, performed using OAuthGuard. Section \ref{sec:discussion} discusses
the limitations and deployment of OAuthGuard. Section \ref{sec:conclusion}
concludes the paper.

\section{Background}
\label{sec:background}

\subsection{OAuth 2.0}
\label{sub:OAuth2}

\begin{figure}[htbp]
 \centering
 \includegraphics[width=0.6\textwidth]{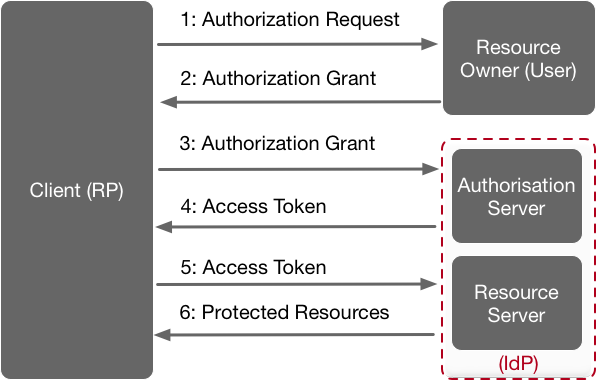}
 \caption{OAuth 2.0 Protocol Flow}
 \label{fig:OAuth2ProcotolFlow}
\end{figure}

The OAuth 2.0 specification \cite{oauth2} describes a system
that allows an application to access resources (typically
personal information) protected by a \emph{resource server} on
behalf of the \emph{resource owner}, through the consumption of
an \emph{access token} issued by an \emph{authorization
server}. In support of this system, the OAuth 2.0 architecture
involves the following four roles (see Fig.
\ref{fig:OAuth2ProcotolFlow}).

\begin{enumerate}

\item The \emph{Resource Owner} is typically an end user.

\item The \emph{Client} is a server which makes requests on behalf of the
    resource owner (the \emph{Client} is the RP when OAuth 2.0 is used for
    SSO).

\item The \emph{Authorization Server} generates access
    tokens for the client, after authenticating the
    resource owner and obtaining its authorization.

\item The \emph{Resource Server} stores the protected resources and
    consumes access tokens provided by an authorization server (this entity
    and the \emph{Authorization Server} jointly constitute the IdP when
    OAuth 2.0 is used for SSO).

\end{enumerate}

Fig.~\ref{fig:OAuth2ProcotolFlow} summarises the OAuth 2.0 protocol. The client
(1) sends an authorization request to the resource owner. In response, the
resource owner generates an authorization grant (or authorization response) in
the form of a \emph{code}, and (2) sends it to the client. After receiving the
authorization grant, the client initiates an access token request by
authenticating itself to the authorization server and presenting the
authorization grant, i.e.\ the code issued by the resource owner (3). The
authorization server issues (4) an access token to the client after
successfully authenticating the client and validating the authorization grant.
The client makes a protected source request by presenting the access token to
the resource server (5). Finally, the resource server sends (6) the protected
resources to the client after validating the access token.

The OAuth 2.0 framework defines four ways for RPs to obtain
access tokens, namely \emph{Authorization Code Grant}, \emph{Implicit Grant},
\emph{Resource Owner Password}, and \emph{Client Credentials Grant}. In this
paper we are only concerned with the \emph{Authorization Code Grant}
and \emph{Implicit Grant} protocol flows. Note that, in the descriptions below, protocol parameters given in bold font are defined as required (i.e.\ mandatory) in the OAuth 2.0 Authorization Framework \cite{oauth2}.

\subsection{OpenID Connect}

OpenID Connect 1.0 \cite{openidconnect} builds an identity layer on top of the
OAuth 2.0 protocol. The added functionality enables RPs to verify an end user
identity by relying on an authentication process performed by an \emph{OpenID
Provider (OP)}\@.  In order to enable an RP to verify the identity of an end
user, OpenID Connect adds a new type of token to OAuth 2.0, namely the
\emph{id\_token}. This complements the access token and code, which are already
part of OAuth 2.0.  An \emph{id\_token} contains claims about the
authentication of an end user by an OP, together with any other claims
requested by the RP\@. OpenID Connect supports three authentication flows
\cite{openidconnect}, i.e.\ ways in which the system can operate, namely
\emph{Hybrid Flow}, \emph{Authorization Code Flow} and \emph{Implicit Flow}.

\subsection{OAuth 2.0 used for SSO}
\label{sub:OAuth2SSO}

In order to use OAuth 2.0 as the basis of an SSO system: the resource server
and authorization server together play the IdP role; the client plays the role
of the RP; and the resource owner corresponds to the user. OAuth 2.0 and OpenID
Connect SSO systems build on user agent (UA) redirections, where a user (U)
wishes to access services protected by the RP which consumes the access token
generated by the IdP\@. The UA is typically a web browser. The IdP provides
ways to authenticate the user, asks the user to grant permission for the RP to
access the user's attributes, and generates an access token on behalf of the
user. After receiving the access token, the RP can access the user's attributes
using the API provided by the IdP\@.

\subsubsection{RP Registration} 
\label{ssub:oauth_resistration}

The RP must register with the IdP before it can use OAuth 2.0, during which the
IdP gathers security-critical information about the RP, including the RP's
redirect URI (\textbf{\textit{redirect\_uri}}),
i.e.\ the URI to which the UA is redirected after the IdP has generated the
authorization response and sent it to the RP via the UA (for convenience, we
also refer to the redirect URI as the \emph{Google sign-in endpoint}). During
registration, the IdP issues the RP with a unique identifier
(\textbf{\textit{client\_id}}) and, optionally, a secret
(\textit{client\_secret}). If defined, \textit{client\_secret} is used by the
IdP to authenticate the RP in the Authorization Code Grant flow.

\subsubsection{Authorization Code Grant} 
\label{ssub:oauthorization_code_flow}


The OAuth 2.0 Authorization Code Grant is very similar to the OpenID Connect
Authorization Code Flow; for simplicity, we only give the description of the
OAuth 2.0 Authorization Code Grant. It relies on information established during
the registration process, as described in Section
\ref{ssub:oauth_resistration}. The protocol proceeds as follows.

\begin{enumerate}
\item U $\rightarrow $ RP: The user clicks a login button on the
RP website, as displayed by the UA, which causes the UA to
send an HTTP request to the RP.

\item RP $\rightarrow$ UA: The RP produces an OAuth 2.0
    authorization request and sends it back to the UA. The
    authorization request includes
    \textbf{\textit{client\_id}}, the identifier for the
    client which the RP registered with the IdP
    previously; \textbf{\textit{response\_type=code}},
    indicating that the Authorization Code Grant method is
    requested; \textit{redirect\_uri}, the URI to which the
    IdP will redirect the UA after access has been granted;
    \textit{state}, an opaque value used by the RP to
    maintain state between the request and the callback
    (step 6 below); and \textit{scope}, the scope of the
    requested permission.

\item UA $\rightarrow$ IdP: The UA redirects the request received in step 2
    to the IdP.

\item IdP $\rightarrow$ UA: The IdP first compares the value of
    \textit{redirect\_uri} it received in step 3 (embedded
    in the authorization request) with the registered value;
    if the comparison fails, the process terminates. If the user has already been
    authenticated by the IdP, then the next step is
    skipped. If not, the IdP returns a login form which is
    used to collect the user authentication information.

\item U $\rightarrow$ UA $\rightarrow$ IdP: The user
    completes the login form and grants permission for the
    RP to access the attributes stored by the IdP.

\item IdP $\rightarrow$ UA $\rightarrow$ RP: After (if
    necessary) using the information provided in the login
    form to authenticate the user, the IdP generates an
    authorization response and redirects the UA back to the
    RP\@. The authorization response contains
    \textbf{\textit{code}}, the authorization code
    (representing the authorization grant) generated by the
    IdP; and \textit{state}, the value sent in step 2.

\item RP $\rightarrow$ IdP: The RP produces an access token
    request and sends it to the IdP token endpoint directly
    (i.e.\ not via the UA)\@. The request includes
    \textbf{\textit{grant\_type=authorization\_code}},
    \textbf{\textit{client\_id}}, \textit{client\_secret}
    (if the RP has been issued one),
    \textbf{\textit{code}} (generated in step 6), and the
    \textbf{\textit{redirect\_uri}}.

\item IdP $\rightarrow$ RP: The IdP checks
    \textbf{\textit{client\_id}}, \textit{client\_secret} (if present),  \textbf{\textit{code}} and
    \textbf{\textit{redirect\_uri}} and, if the checks succeed, responds to the RP
    with \textit{access\_token}.

\item RP $\rightarrow$ IdP: The RP passes
    \textit{access\_token} to the IdP via a defined API to
    request the user attributes.

\item IdP $\rightarrow$ RP: The IdP checks
    \emph{access\_token} (how this works is not specified in the OAuth 2.0 specification) and, if
    satisfied, sends the requested user attributes to the
    RP\@.
\end{enumerate}

\subsubsection{Implicit Grant}
\label{ssub:implicit_grant}
The OAuth 2.0 Implicit Grant is very similar to the OpenID Connect Implicit Flow and Hybrid Flow; for simplicity, we only give the description of the OAuth 2.0 Implicit Grant. This flow has a similar sequence of
steps to Authorization Code Grant. We specify below only those
steps where the Implicit Grant flow differs from the
Authorization Code Grant flow.

\begin{enumerate}

\item[2.] RP $\rightarrow$ UA: The RP produces an OAuth 2.0
    authorization request and sends it back to the UA\@.
    The authorization request includes
    \textbf{\textit{client\_id}}, the identifier for the
    client which the RP registered with the IdP
    previously;  \textbf{\textit{response\_type=token}},
    indicating that the Implicit Grant  is requested;
    \textit{redirect\_uri}, the URI to which the IdP will
    redirect the UA after access has been granted;
    \textit{state}, an opaque value used by the RP to
    maintain state between the request and the callback
    (step 6 below); and \textit{scope}, the scope of the
    requested permission.

\item[6.] IdP $\rightarrow$ UA $\rightarrow$ RP: After (if
    necessary) using the information provided in the login
    form to authenticate the user, the IdP generates an
    access token and redirects the UA back to the RP using the value of
   \textit{redirect\_uri} provided in step 2. The access token
    is appended to \textit{redirect\_uri} as a URI fragment
    (i.e.\ as a suffix to the URI following a \# symbol).

\end{enumerate}

As URI fragments are not sent in HTTP requests, the access token is
not immediately transferred when the UA is redirected to the
RP\@. Instead, the RP returns a web page (typically an HTML
document with an embedded script) capable of accessing the full
redirection URI, including the fragment retained by the UA, and
extracting the access token (and other parameters) contained in
the fragment; the retrieved access token is returned to the
RP\@. The RP can now use this access token to retrieve data
stored at the IdP\@.

\section{Analysing the security of OAuth 2.0}
\label{sec:related_work}

OAuth 2.0 has been analysed using formal methods. Pai et al.\ \cite{pai11}
confirmed a security issue described in the OAuth 2.0 Threat Model
\cite{oauth2threat} using the Alloy Framework \cite{alloy}. Chari et al.\
analysed OAuth 2.0 in the Universal Composability Security framework
\cite{DBLP:journals/iacr/ChariJR11} and showed that OAuth 2.0 is secure if all
the communications links are SSL-protected. Frostig and Slack \cite{frostig11}
discovered a cross-site request forgery attack in the Implicit Grant flow of
OAuth 2.0, using the Murphi framework \cite{DBLP:conf/cav/Dill96}. Bansal et
al.\ \cite{DBLP:journals/jcs/BansalBDM14} analysed the security of OAuth 2.0
using the WebSpi \cite{WebSpi} and ProVerif models \cite{ProVerif}. Fett et
al.\ \cite{fett2017web} performed a formal security analysis of OpenID Connect.
However, all this work is based on abstract models, and so delicate
implementation details are ignored.

The security properties of real-world OAuth 2.0 implementations
have also been examined. Wang et al.\
\cite{DBLP:conf/sp/WangCW12} examined deployed SSO systems,
focusing on a logic flaw present in many such systems,
including OpenID\@. In parallel, Sun and Beznosov
\cite{DBLP:conf/ccs/SunB12} also studied deployed OAuth 2.0
systems. Later, Li and Mitchell \cite{DBLP:conf/isw/LiM14}
examined the security of deployed OAuth 2.0 systems providing
services in Chinese. In parallel, Zhou and Evans
\cite{DBLP:conf/uss/ZhouE14} conducted a large scale study of
the security of Facebook's OAuth 2.0 implementation. Chen et
al.\ \cite{DBLP:conf/ccs/ChenPCTKT14}, and Shehab and Mohsen
\cite{DBLP:conf/codaspy/ShehabM14} have looked at the security
of OAuth 2.0 implementations on mobile platforms. Li and
Mitchell \cite{wanpeng:dimva2016} conducted an empirical study
of the security of the OpenID Connect-based SSO service
provided by Google.

Fett et al.\ \cite{fett2016comprehensive} proposed an IdP Mix-Up attack against
RPs that support multiple IdPs. In their attack, a network attack is needed to
modify the http or https messages generated by the RP in step 1 (see Section
\ref{ssub:oauthorization_code_flow}). Li and Mitchell \cite{wanpeng2016does}
argued that the IdP Mix-Up attack would not be a genuine threat to the security
of OAuth 2.0 if IdP implementations were strictly following the standard. Li and Mitchell \cite{wanpeng:spw2018}  proposed a Partial Redirection URI Manipulation attack against RPs that support multiple IdPs. A 2016 study conducted by Yang et al.\ \cite{DBLP:conf/ccs/YangLLZH16} revealed
that 61\% of 405 websites using OAuth 2.0 (chosen from the 500 top-ranked US
and Chinese websites) did not implement CSRF countermeasures; even worse, for
those RPs which support the state parameter, 55\% of them are still vulnerable
to CSRF attacks because of misuse/mishandling of the state parameter. They also
disclosed four scenarios where the state parameter can be misused by RP
developers.  Most recently, Yang, Lau and Shi \cite{DBLP:conf/acns/YangLS17}
conducted a large scale study of Android OAuth 2.0-based SSO systems. They
found three previously unknown security flaws among first-tier identity
providers and a large number of popular third party apps.

These practical studies suggest that in practice many real-world OAuth 2.0 and
OpenID Connect systems  contain security vulnerabilities, often because
of implementation errors made by RP developers. In some cases these errors
result from a lack of clear guidance from IdPs. Regardless of the causes,
these vulnerabilities pose a significant threat to end users, and addressing
this threat has motivated the work described in this paper.


In recent work conducted in parallel to that described here\footnote{The source
code of OAuthGuard\footnote{\url{https://github.com/wanpengli/OAuthGuard}} was
first made available at \url{github.com} in February 2018.}, Calzavara et al.\
\cite{calzavara2018wpse} proposed WSPE, a web browser security monitor for
OAuth 2.0. OAuthGuard and WSPE have some similar functionalities, e.g.\ being
able to detect some common OAuth 2.0 attacks and provide mitigations for the
user (see Table \ref{table:comparison}). One major advantage of OAuthGuard is
that it is able to detect and provide protections for five common
vulnerabilities for users, whereas WSPE can only detect three.  A more detailed
comparison of OAuthGuard with WPSE is provided in Section \ref{sec:discussion}.

\section{Vulnerabilities} 
\label{sec:vulnerabilities}

The design of OAuthGuard was motivated by the work of Li and Mitchell
\cite{wanpeng:dimva2016} and Yang et al.\ \cite{DBLP:conf/ccs/YangLLZH16}. They
examined the security of real-world OAuth 2.0 and OpenID Connect
implementations and identified a range of vulnerabilities; they also proposed
mitigations designed to enable RPs to make their OAuth 2.0 and OpenID connect
systems secure.  However, none of these mitigations help protect users who are
employing an insecure OAuth 2.0 or OpenID Connect implementation.  OAuthGuard
is intended to help meet this need.

\label{ssub:vulnerabilities_detect}


OAuthGuard can detect five classes of OAuth 2.0 or OpenID Connect
vulnerabilities --- four of these vulnerabilities have previously been
discussed (see, for example, Li and Mitchell, \cite{wanpeng:dimva2016}) --- the
only vulnerability not previously discussed is the `privacy leak' issue, i.e.\
the fifth in the list below. Impersonation attacks only apply to the Implicit
Grant flow, as described in Section \ref{ssub:implicit_grant}; the other four
attacks affect both flows, as defined in Section \ref{sub:OAuth2SSO}.

\begin{itemize}
  \item \textbf{CSRF Attack Threat Detection}. CSRF attacks against the OAuth 2.0
      \emph{redirect\_uri} \cite{oauth2threat} can allow an attacker to
      obtain authorization to access OAuth-protected resources without the
      consent of the user. Such an attack is possible for both the
      Authorization Code Grant Flow and the Implicit Grant Flow.

  One possible CSRF attack involves an attacker engaging with the target RP
  using its own device, and acquiring a \textit{code}, \accesstoken or
  \idtoken for the attacker's own resources. The attacker then aborts the
  redirect flow back to the RP, and, by means of a CSRF, instead causes the
  victim user to send the (attacker's) redirect flow back to the target
  RP\@. The target RP receives the redirect, fetches the (attacker's)
  attributes from the IdP, and associates the victim user's RP session with
  the attacker's resources accessed using the tokens. The victim user then
  accesses resources on behalf of the attacker.  The impact of such an
  attack depends on the resources accessed. For example, the user might
  upload private data to the RP, thinking it is uploading information to
  its own profile, and this data will subsequently be available to the
  attacker. Alternatively, as described by Li and Mitchell
  \cite{DBLP:conf/isw/LiM14}, an attacker can use a CSRF attack to control
  a victim user's RP account without knowing the user's username and
  password.

  A 2016 study conducted by Yang et al.\
  \cite{DBLP:conf/ccs/YangLLZH16} revealed that 61\% of 405
  websites using OAuth 2.0 (chosen from the 500 top-ranked
  US and Chinese websites) did not implement the `standard'
  CSRF countermeasures, notably including use of the state
  parameter; even worse, of those RPs which did support the
  state parameter, 55\% were still vulnerable to CSRF
  attacks because of incorrect use of this parameter. They
  also described four scenarios in which the state
  parameter can be misused by RP developers. Given these
  variations in incorrect implementations, it is difficult
  to devise a universally applicable method to
  automatically detect a CSRF attack threat. As discussed
  in greater detail in the next section, the technique
  OAuthGuard uses to detect this threat is simply to check
  whether a \textit{state} parameter is present in an OAuth
  2.0 response. If no such parameter is present, then
  OAuthGuard reports that the RP is vulnerable to a CSRF
  attack. Thus OAuthGuard is not able to detect all RPs
  that are vulnerable to CSRF attacks, e.g.\ arising from
  incorrect use of the parameter.

  \item \textbf{Impersonation attacks}. This vulnerability stems from
      confusion about authentication and authorization. In OAuth 2.0, an
      \accesstoken is intended for authorization purposes, and it is not
      tied to any specific RP\@.  As the \accesstoken is a bearer token, it
      can be used by any RP that gains access to it. If an RP submits only
      an \accesstoken to their Google sign-in endpoint, a malicious RP can
      submit a victim user's \accesstoken, issued to the malicious RP by
      Google, to the RP's Google sign-in endpoint. The RP can use this
      \accesstoken to get victim user information from Google, and then get
      full access to the victim user's account at the RP\@.

  \item \textbf{Authorization Flow Misuse}. As described in Section
      \ref{sec:background}, OAuth 2.0 has four authorization flows and
      OpenID Connect has three authentication flows. RP developers must
      choose an appropriate flow and implement the OAuth 2.0 or OpenID
      Connect protocol correctly. According to the OAuth 2.0 and OpenID
      Connect standards, only a \textit{code} should be submitted back to
      the RP's Google sign-in endpoint as evidence that the user has been
      authenticated. However, in reality, many RPs submit a combination of
      \textit{code}, \accesstoken and \idtoken back to their Google sign-in
      endpoint.  As discussed by Li and Mitchell \cite{wanpeng:dimva2016},
      this can lead to serious vulnerabilities.

  \item \textbf{Unsafe Token Transfers}. The main purpose of OAuth 2.0 and
      OpenID Connect is to allow an RP to access user information stored at
      an IdP without giving the RP the user's credentials for the IdP\@.
      This is achieved using a \textit{code}, \accesstoken or \idtoken.
      These tokens are vitally important, and hence they need to be
      protected when transferred between the RP and Google (e.g.\ using
      HTTPS)\@. However, as has been discussed by Li and Mitchell
      \cite{wanpeng:dimva2016}, many RPs do not use HTTPS to protect the
      Google sign-in data transfers.

  \item \textbf{Privacy Leaks}. When a user uses the Google service to
      authenticate to an RP website, the user's \textit{code}, \accesstoken
      or \idtoken, retrieved by the RP from Google, should not be revealed
      to any other parties. We consider two cases where such a token may be
      revealed to a third party, which we refer to as a \emph{privacy
      leak}; we further distinguish between \textit{intentional privacy leaks}
      and \textit{referer (unintentional) privacy leaks}, depending on whether the RP is
      aware of the leak or not. An unintentional privacy leak might occur
      when an RP includes third party content in its Google sign-in
      endpoint; an intentional privacy leak occurs when an RP deliberately
      sends user tokens to a third party.

\end{itemize}

\section{OAuthGuard} 
\label{sec:oauthguard}

\begin{figure}[htbp]
 \centering
 \includegraphics[width= 0.8\textwidth]{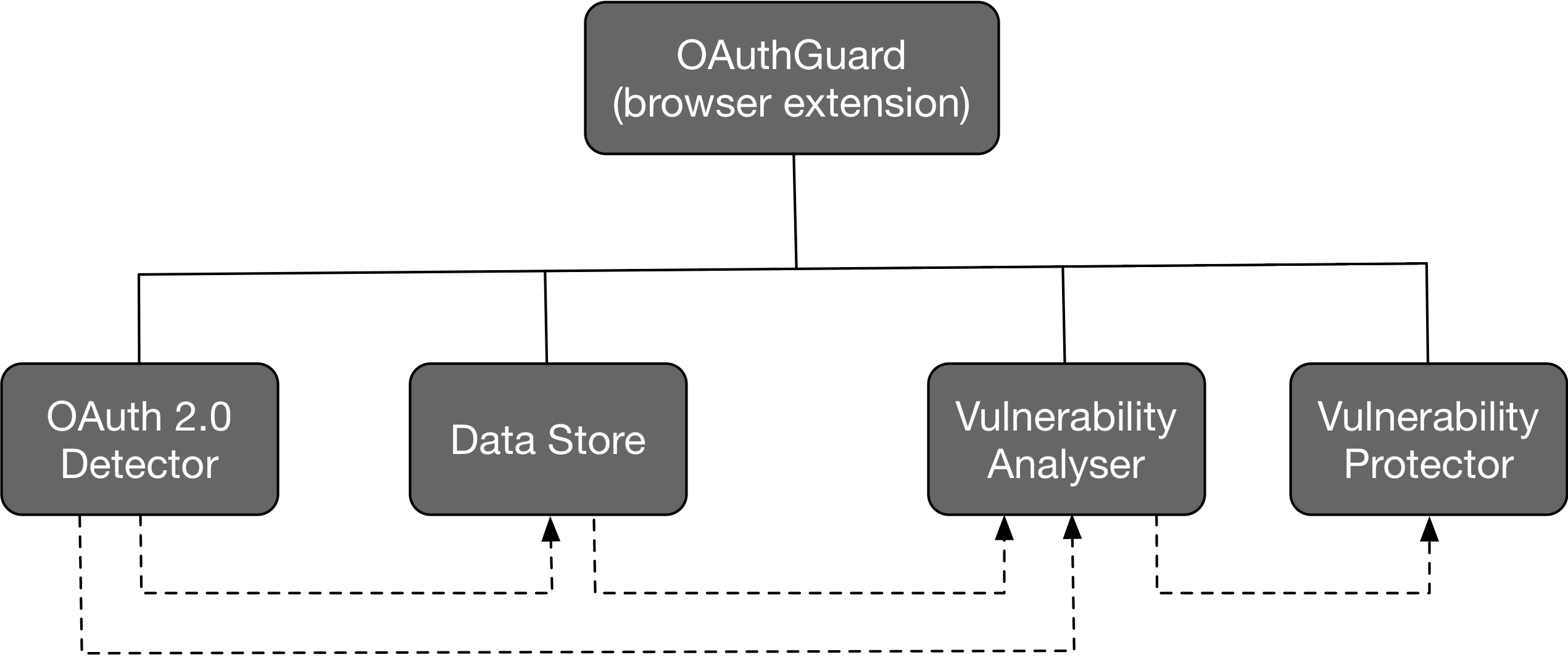}
 \caption{OAuthGuard Components}
 \label{fig:oauthguard}
\end{figure}

OAuthGuard, a JavaScript Chrome browser extension, which is freely available via the Chrome web
store\footnote{\url{https://chrome.google.com/webstore/detail/oauthguard/phamalogfapdjegegmghgcihhpabocfn}}, contains three main
components: the \textit{OAuth 2.0 Detector}, the \textit{Vulnerability
Analyser}, and the \textit{Vulnerability Protector} (see Fig.\
\ref{fig:oauthguard}). The OAuth 2.0 Detector monitors every HTTP request and
extracts the OAuth 2.0 request or response metadata (see Listing
\ref{listing:oauth2metadata} in Appendix) if the request is an OAuth 2.0 request or
response. The Vulnerability Analyser analyses the OAuth 2.0 request and
response reported by the OAuth 2.0 Detector, with the goal of identifying the
possible vulnerabilities described in Section
\ref{ssub:vulnerabilities_detect}. Once a vulnerability has been detected by
the Vulnerability Analyser, the Vulnerability Protector is triggered and
appropriate mitigations are executed.


\subsection{Vulnerability Mitigation} 
\label{ssub:vulnerabilities_protection}

OAuthGuard protects against all five of the vulnerabilities in Section
\ref{ssub:vulnerabilities_detect}. We next describe how OAuthGuard mitigates
these vulnerabilities.

\begin{itemize}
  \item \textbf{CSRF Attack Protection}.
OAuthGuard is designed to mitigate CSRF attacks even if the RP does not
      implement any countermeasures against such attacks (e.g.\ if it does
      not include a state parameter in the authorization response). To
      achieve this, OAuthGuard uses the CSRF countermeasures recently
      proposed by Li and Mitchell \cite{wanpeng:pst2018}. The idea is
      that, when used correctly, the referer header of the OAuth 2.0
      response should point to either the RP domain or the IdP domain; this
      can be used to detect CSRF attacks. However, one limitation of this
      approach is that, if the \redirecturi of the RP uses HTTP, the
      referer header will be suppressed by the user agent \cite{http}
      (i.e.\ the necessary domain information will be removed). Thus,
      OAuthGuard can only be used to mitigate CSRF attacks for RPs that use
      HTTPS to transfer their OAuth 2.0 response.

      To implement this mitigation, OAuthGuard first checks to see whether
      HTTPS has been used to transfer the OAuth 2.0 response; if not, it
      simply ignores the OAuth 2.0 response for compatibility reasons (so
      as not to block RPs using HTTP to transfer their OAuth 2.0 response).
      That is, OAuthGuard accepts all requests for RPs that use HTTP to
      deliver their OAuth 2.0 response. Otherwise, i.e.\ if HTTPS is used,
      it checks whether the HTTP referer header of the OAuth 2.0 response
      points to either the Google domain or the RP's domain; if not then
      OAuthGuard knows it is a CSRF attack against the RP's Google sign-in
      endpoint, drops the message, and notifies the user that it has
      blocked a CSRF attack attempt.

      This technique works with most RPs, although a few RPs use a proxy
      service (e.g.\ gigya) to implement support for Google sign-in, or use
      a domain other than the domain registered with Google as their Google
      sign-in endpoint.  We whitelisted these RP domains so that, in such
      cases, OAuthGuard will not block the OAuth 2.0 response. In summary,
      OAuthGuard implements strict Referer validation
      \cite{DBLP:conf/ccs/BarthJM08} to protect against CSRF attacks for
      RPs that use HTTPS to deliver their OAuth 2.0 response; that is,
      OAuthGuard blocks all HTTPS requests whose Referer header has an
      incorrect value (e.g.\ an empty referer header).

  \item \textbf{Impersonation Attack Warning}. OAuthGuard is able to
      discover the RP's Google sign-in endpoint, and can also extract all
      three types of token from an OAuth 2.0 Response HTTP message. If only
      an \accesstoken is submitted to the PR's Google sign-in endpoint,
      then OAuthGuard notifies the user that the RP's website might be
      vulnerable to an impersonation attack, and that the user is
      recommended to stop using Google sign-in with that RP\@.

  \item \textbf{Authorization Flow Misuse Warning}.  OAuthGuard is able to
      detect an Authorization Flow Misuse vulnerability, as described in
      Section \ref{ssub:vulnerabilities_detect}.  However, it cannot
      determine which token is used by the RP to authenticate the user. As
      a result, OAuthGuard does not implement any active mitigations, but
      simply generates a warning message to the user.

  \item \textbf{Unsafe Token Transfer Protection}. OAuthGuard is able to
      extract the protocol message used to transfer an OAuth 2.0 response.
      If HTTP is used, OAuthGuard attempts to redirect the response using
      HTTPS before the response leaves the user's browser. Of course, this
      measure only works if HTTPS is available at the RP\@.

  \item \textbf{Privacy Leak Protection}. As discussed in Section
      \ref{ssub:vulnerabilities_detect}, if either of the two types of
      privacy leak is detected, OAuthGuard blocks the transfers and
      notifies the users that it has blocked an attempted privacy leak.
      Another possible mitigation would be to remove the tokens from the
      referer header instead of blocking the entire request. We chose to
      block the request because we want to discourage users from using the
      Google sign-in service for an RP that leaks user data to a third
      party.

\end{itemize}

\subsection{OAuth 2.0 Detector} 
\label{sub:oauth_2_0_detector}

\begin{figure}[htbp]
 \centering
 \includegraphics[width= 0.8\textwidth]{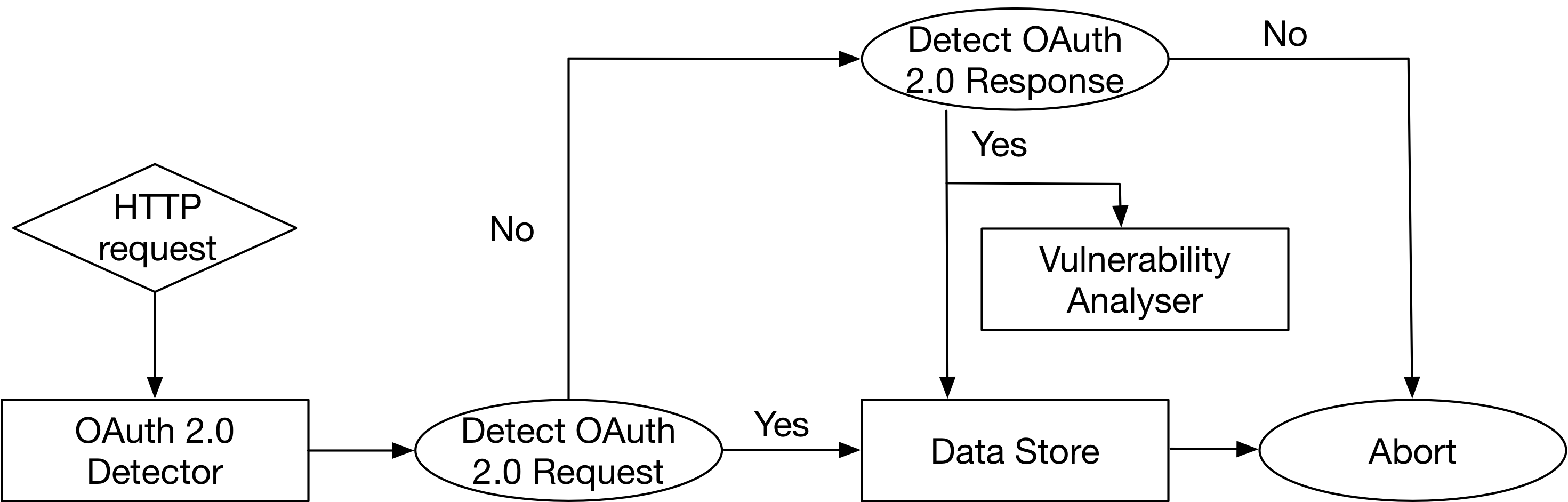}
 \caption{OAuthGuard OAuth 2.0 Detector Overview}
 \label{fig:detectorflow}
\end{figure}

Figure \ref{fig:detectorflow} shows the workflow of the OAuth 2.0 Detector. The
OAuth Detector first examines every received HTTP request to check whether it is an
OAuth 2.0 request; this involves scanning the url of the request for the
keywords \textit{oauth} and \textit{redirect\_uri}.  If both keywords are
present, the HTTP request is deemed to be an OAuth 2.0 request; the OAuth 2.0
Detector then extracts the OAuth 2.0 request metadata (see Listing
\ref{listing:oauth2metadata} in Appendix) from the HTTP request and saves this metadata to
the extension's
localStorage\footnote{\url{https://developer.mozilla.org/en-US/docs/Web/API/Web_Storage_API#localStorage}}
using RPDomain as key.

Otherwise, i.e.\ if these keywords are not both present, the OAuth 2.0 Detector
scans the HTTP request for a \textit{code}, \accesstoken or \idtoken; if one of
these tokens is identified, the HTTP request is deemed to be an OAuth 2.0
response.  In this case the OAuth 2.0 Detector extracts the OAuth 2.0 response
metadata from the HTTP request and saves it to localStorage using RPDomain as
key.

\subsection{Vulnerability Analyser} 
\label{sub:vulnerability_analyser}

\begin{figure}[htbp]
 \centering
 \includegraphics[width= 0.8\textwidth]{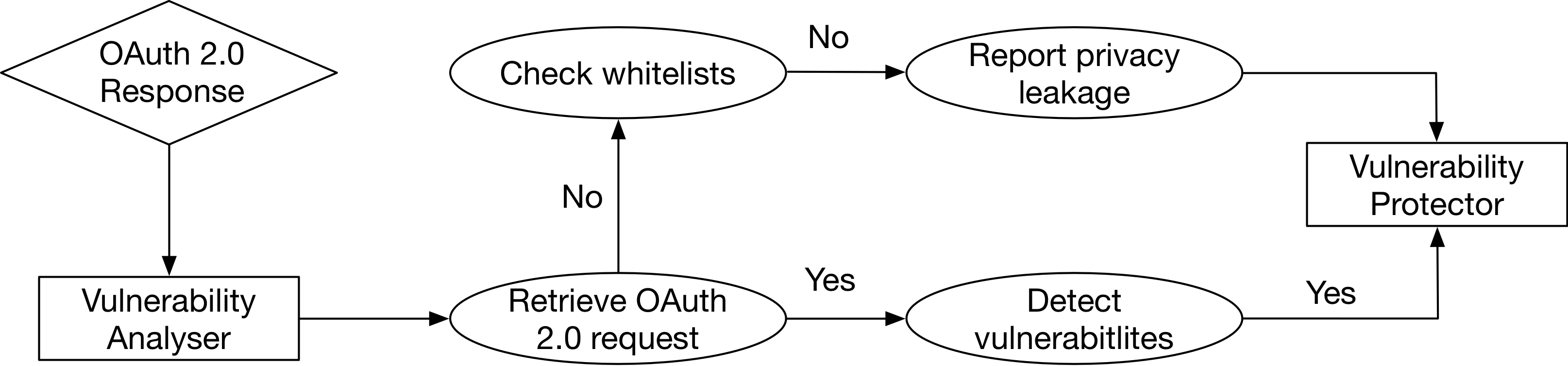}
 \caption{OAuthGuard Vulnerability Analyser Overview}
 \label{fig:analyserflow}
\end{figure}

Figure \ref{fig:analyserflow} shows the workflow of the Vulnerability Analyser.
Whenever an OAuth 2.0 response it reported by the OAuth 2.0 Detector, the
Vulnerability Analyser is triggered. It first retrieves the OAuth 2.0 request
(if any) using the RPDomain from the OAuth 2.0 response.
\begin{itemize}
\item If no OAuth 2.0 request is retrieved, then the OAuth 2.0 response
    might be from an RP that is either using a proxy service (e.g.\ gigya)
    to implement Google sign-in, or a domain other than the domain
    registered with Google as its Google sign-in endpoint. Such RPs and
    proxy services are whitelisted in OAuthGuard. If neither of these two
    cases applies, i.e.\ the domain is not in the whitelist, the OAuth 2.0
    response is deemed to be an intentional privacy leak.
\item If an OAuth 2.0 request is retrieved, the Vulnerability Analyser uses
    the OAuth 2.0 request and response to identify possible vulnerabilities
    as follows.
    \begin{enumerate}
     \item \textbf{Detection of CSRF Threats}. If a \textit{state} parameter is not present in an
         OAuth 2.0 response, then
         OAuthGuard reports that the RP is vulnerable to a CSRF attack.
    \item \textbf{Detection of an Impersonation attack}. If only an
        \accesstoken is detected in the OAuth 2.0 response, OAuthGard
        reports a possible Impersonation attack.
    \item \textbf{Detection of Authorization Flow Misuse}. If a
        combination of \textit{code}, \accesstoken and \idtoken is
        detected in the OAuth 2.0 response, OAuthGuard reports an
        Authorization Flow Misuse.
    \item \textbf{Detection of Unsafe Token Transfer}. OAuthGuard
        checks whether the RP is using HTTP or HTTPS to transfer the
        OAuth 2.0 response. If HTTP is detected, it reports an Unsafe
        Token Transfer threat.
    \item \textbf{Detection of Privacy Leaks}. OAuthGuard uses a
        specific Referer Leakage Detection module to detect
        Unintentional Privacy Leak vulnerabilities. This module first
        looks for a \textit{code}, \accesstoken or \idtoken in a
        referer header (if present in an HTTP request). If any of these
        tokens are identified, the module extracts the domains of the
        referer header and the HTTP request, and checks whether they
        are the same. If not, it reports that an Unintentional Privacy
        Leak vulnerability has been detected.
    \end{enumerate}
\end{itemize}

\subsection{Vulnerability Protector} 
\label{sub:vulnerability_protector}


Depending on which type of vulnerability has been reported by the Vulnerability
Analyser, the Vulnerability Protector executes the following actions:
\begin{itemize}
\item it blocks an HTTP request if a Privacy Leak vulnerability is
    detected;
\item it tries to redirect an OAuth 2.0 response using HTTPS if an Unsafe
    Token Transfer vulnerability is detected;
\item it warns the user if the RP is vulnerable to an Impersonation attack;
\item it blocks an OAuth 2.0 response if a CSRF attack is detected.

\end{itemize}

\section{A Case Study}
\label{sec:case_study}

We used OAuthGuard to help understand the degree to which RPs using the Google
OAuth service are vulnerable to known threats.  This involved manually running
OAuthGuard against the top-ranked 1,000 websites from
majestic.com\footnote{\url{https://majestic.com/reports/majestic-million}} as
of 12 December 2017. 137 of these 1,000 websites support Google sign-in. We
used a Macbook Pro (late 2013) running macOS High Sierra 10.13.1 and Chrome
browser version 63.0.3239.132.

As discussed earlier, OAuthGuard detects CSRF attack threats by checking
whether a state parameter is present in an OAuth 2.0 response.  To supplement
the automated threat detection, we also manually looked through all the RPs for
which a CSRF threat was not reported by OAuthGuard to discover whether these
RPs are actually vulnerable to a CSRF attack. While OAuthGuard reported CSRF
attack threats for 28 of the 137 RPs, we manually identified a further 25 that
are vulnerable to CSRF attacks.

\label{sub:study_result}

\begin{figure}[htbp]
 \centering
 \includegraphics[width= 0.8\textwidth]{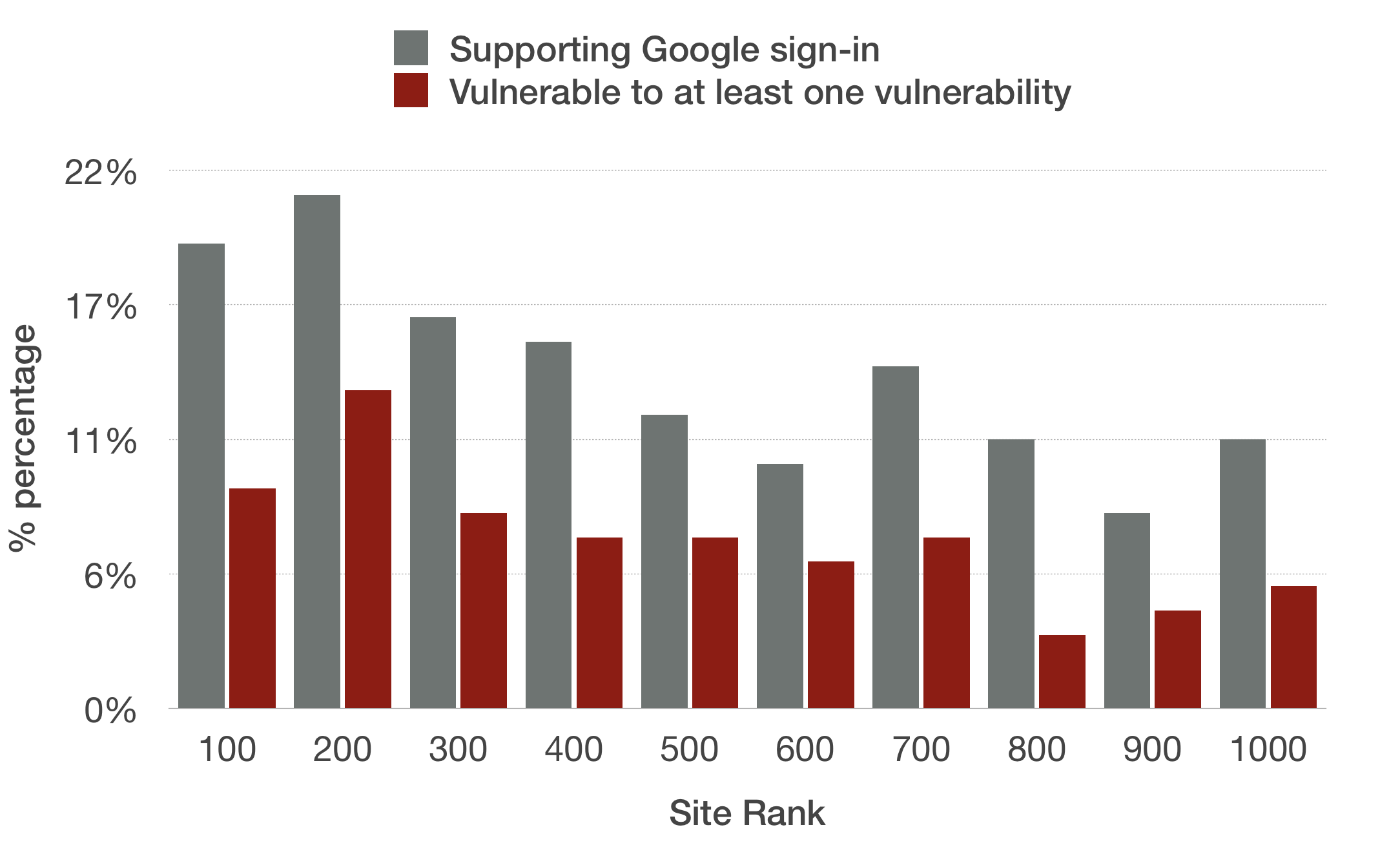}
 \caption{Google Sign-in integration results by site rank}
 \label{fig:trend}
\end{figure}

Figure \ref{fig:trend} divides the 1,000 sites we examined into groups of 100
(starting with those ranked highest), and for each group of 100 indicates (a)
the percentage supporting Google SSO, and (b) of those that do support Google
SSO which were found to possess at least one vulnerability.  The graph suggests
that the more popular sites are a little more likely to support Google sign-in
and also slightly more likely to possess implementation vulnerabilities. Whilst
the former result is not surprising, the latter is somewhat alarming, since one
might expect popular sites to have more resources to devote to ensuring site
security.

Unsurprisingly, we got similar results to those of the 2016
Yang et al.\ study \cite{DBLP:conf/ccs/YangLLZH16}. We can
summarise our findings as follows (noting that in each case
they apply to the 137 RPs that support Google SSO).
\begin{itemize}
\item 53 RPs (39\%) are vulnerable to a CSRF attack against their OAuth 2.0
    \redirecturi endpoint;
\item 21 RPs (15\%) misuse authorization flows, of which 13 are vulnerable
    to an impersonation attack;
\item 9 RPs leak tokens though referer headers; of these, two explicitly
    send user tokens to third party websites;
\item 13 RPs did not implement https to protect the transfer of user
    tokens.
\item A total of 69 RPs (50\%) possessed at least one vulnerability of the
    types discussed in Section \ref{ssub:vulnerabilities_detect}.
\end{itemize}

\label{ssub:examples}

To illustrate the potential risks in real-world websites, we next give an
example of vulnerabilities detected by OAuthGuard. Ranked 61st on the list,
\url{issuu.com} is the world's largest electronic publishing platform.
OAuthGuard detected that \url{issuu.com} is vulnerable to Referer token leaks,
CSRF attacks, as well as Impersonation attacks. After the user has used Google
sign-in to log in to \url{issuu.com}, it sends an \accesstoken directly to its
Google sign-in endpoint without implementing any CSRF countermeasures.
Moreover, its Google sign-in endpoint contains content (e.g.\ gif and
JavaScript files) from third-party websites, including \url{optimizely.com},
\url{bing.com}, \url{licdn.com}, and \url{quantserver.com}.  When the browser
retrieves this content, it sends the \accesstoken in the Referer header to
these third-party sites. The permissions \url{issuu.com} requests from Google
include access to the user’s profile information and email address, so that
possession of the \accesstoken gives access to this information without user
consent.

\section{Discussion} 
\label{sec:discussion}

\subsubsection{Implementation challenges} 
\label{sub:user_security_and_privacy_protect}

The main design goal of OAuthGuard is to protect user security and privacy when
using Google sign-in. Since each RP implements its own Google sign-in endpoint,
it is hard to devise a solution that will work for all RPs. We next describe
some of the difficulties we encountered in designing OAuthGuard, and the
trade-offs we made to enable it to operate.

\emph{CSRF Protection} 
\label{ssub:csrf_protection} As discussed in Section
\ref{ssub:vulnerabilities_protection}, OAuthGuard implements strict referer
validation \cite{DBLP:conf/ccs/BarthJM08} to protect users against CSRF attacks
for RPs using HTTPS to deliver an OAuth 2.0 response. This works as in this
case the referer header domain should be either the IdP's domain or the RP's
domain. However, some RPs use a proxy service (e.g.\ gigya) to implement Google
sign-in, or use a domain other than the domain registered with Google as their
Google sign-in endpoint. For example, \url{chicagotribune.com} registers
\url{https://signin.chicagotribune.com/GS/GSLogin.aspx?} as its Google sign-in
\redirecturi, but uses the domain
\url{https://ssor.tribdss.com/assets/sso_popup.html} to display its Google
sign-in button. If no other checks were implemented, OAuthGuard would
incorrectly report a CSRF attack on \url{chicagotribune.com}, as the referer
header domain \url{tribdss.com} does not equal either \url{google.com} or
\url{chicagotribune.com}. In order to make OAuthGuard compatible with such RPs,
we chose to whitelist all such domain names in the OAuthGuard source code. In
total we whitelisted 11 domains (8\%) from the set of 137 RPs. Even given these
difficulties, OAuthGuard can protect user security for 48 of the 53 RPs (91\%)
which we found to be vulnerable to CSRF attacks. The other five RPs use HTTP to
deliver the OAuth 2.0 response, and as a result the OAuthGuard CSRF
countermeasure does not work.

\emph{Privacy Protection} \label{ssub:privacy_protection} As described in
Section \ref{sub:study_result}, OAuthGuard identified nine RPs that leak user
tokens to third party websites, either intentionally or unintentionally.
OAuthGuard blocked all the token-leaking HTTP requests for these nine RPs. In
total, OAuthGuard blocked 75 HTTP requests that leak user tokens for these nine
RPs. Blocking third party requests that leak tokens might prevent users from
using Google sign-in to log in to the relevant RPs\@. However, we decided to
block these requests as it will discourage users from using insecure Google
sign-in implementations; most importantly it prevents unauthorised token
disclosure, which could have a serious negative impact on user privacy.

\emph{Impersonation Attack Warnings} \label{ssub:impersonation_warning} It is
up to the RP to decide which types of tokens it should submit back to its
Google sign-in endpoint.  If tokens are used inappropriately, the only thing
OAuthGuard can do is to warn users that an RP is vulnerable to an impersonation
attack, and suggest that users should not employ Google sign-in at these RPs.

\emph{HTTPS Upgrade} \label{ssub:https_upgrade} If OAuthGuard detects an OAuth
2.0 response transferred using HTTP, it attempts to redirect it using HTTPS\@.
Of course, this protection only works with RPs that implement HTTPS on their
website. In our study, OAuthGuard was able to upgrade the protocol to HTTPS for
8 of the 13 RPs (62\%) that use HTTP to transfer an OAuth 2.0 response (in each
case the HTTPS upgrade resulted in a successful login). For RPs not supporting
HTTPS, OAuthGuard will by default make the Google sign-in service unavailable;
to give the user flexibility in which sites they are able to use, OAuthGuard
enables users to turn off the HTTPS upgrade function.

\subsubsection{Comparison with WPSE}

The OAuthGuard approach to protecting against CSRF attacks is more efficient
that employed by WSPE, because WSPE blocks any OAuth 2.0 response which
does not contain a state parameter (see \cite{calzavara2018wpse} Section
4.1.1); Yang et al.\ \cite{DBLP:conf/ccs/YangLLZH16} found that 61\% of 405
websites using OAuth 2.0 (chosen from the 500 top-ranked US and Chinese
websites) did not implement CSRF countermeasures; this means that WSPE would
block the OAuth 2.0 response for these 61\% websites. Our approach to
mitigating CSRF attacks is to use the CSRF countermeasures recently proposed by
Li and Mitchell \cite{wanpeng:pst2018}.  These build on the observation that,
when used correctly, the referer header of the OAuth 2.0 response should point
to either the RP domain or the IdP domain; this can be used to detect CSRF
attacks. Using this approach, OAuthGuard can protect users against CSRF attacks
even when RPs do not implement any CSRF countermeasures (including the 61\% of
RPs in Yang's study \cite{DBLP:conf/ccs/YangLLZH16}).

\begin{table}[]
\centering
\begin{tabular}{|l|c|c|}
\hline
\multicolumn{1}{|c|}{}    & OAuthGuard & WSPE \\ \hline
CSRF Attacks              & x          & x    \\ \hline
Impersonation Attacks     & x          &      \\ \hline
Privacy Leaks             & x          & x    \\ \hline
Authorization Flow Misuse & x          &      \\ \hline
Unsafe Token Transfers    & x          &      \\ \hline
Mix-IdP Attack            &            & x    \\ \hline
\end{tabular}
\caption{Comparison between OAuthGuard and WSPE} \label{table:comparison}
\end{table}

\subsubsection{Limitations} 
\label{sub:Limitations}

OAuthGuard detects vulnerabilities by analysing HTTP messages.  However, this
approach cannot be used to detect vulnerabilities that can only be found by
deep server-side application scanning. For example, the IdP Mix-Up attack
revealed by Fett et al.\ \cite{fett2016comprehensive} could be detected by RP
developers using program analysis techniques, but cannot be detected by an
external tool with no awareness of the site's implementation details or
internal state. Also, since OAuthGuard blocks HTTP messages that leak user
tokens to third-party websites, it could make the Google sign-in service
unavailable for some RPs.

\subsubsection{Disclosure} 
\label{sub:Reporting}

We reported our findings to seven RPs that are vulnerable to the impersonation
attacks described in Section \ref{ssub:vulnerabilities_detect}. We contacted
them either by email or by submitting a website form. The responses were
disappointing, especially compared with previous experience in reporting
SDK-level vulnerabilities to IdPs \cite{wanpeng:dimva2016}, who typically
responded quickly to vulnerability reports. The lack of response is perhaps
explained by the fact that the vulnerabilities we identified are primarily in
consumer-oriented RP sites, who may not have dedicated security teams or ways
of effectively addressing security issues. So far we have only received
responses from two RPs in which they acknowledge our reports and are working on
fixing the vulnerability; of course, we may receive more responses in the
future --- we certainly hope so.

\subsubsection{Testing and Deployment} 
\label{sub:Deployment}

OAuthGuard has been informally tested by the authors and their colleagues; no
significant usability issues have so far been detected.  Of course, this is
hardly a thorough test, consistent with the fact that OAuthGuard is primarily
intended as a prototype and proof-of-concept. If it is to be very widely
deployed, then further development work will be required to ensure that the
whitelist is expanded to cover all well-used sites that would otherwise fail
the checks. Nonetheless, our informal tests reveal that OAuthGuard as it is
offers an enhanced level of user security and privacy protection.

OAuthGuard is freely available via the Chrome web
store\footnote{\url{https://chrome.google.com/webstore/detail/oauthguard/phamalogfapdjegegmghgcihhpabocfn}},
and the source code is available at
github\footnote{\url{https://github.com/wanpengli/OAuthGuard}}. We hope that
researchers and developers can help to further develop the tool, as well as
enabling support for other OAuth 2.0 systems, such as those of Facebook and
Microsoft.

Apart from end user deployment, OAuthGuard can also be used by RP developers to
check Google sign-in implementations. After the usual development testing, and
before launching support for Google sign in, developers could usefully run
OAuthGuard to detect any residual vulnerabilities.

\section{Conclusion} 
\label{sec:conclusion}

We have described OAuthGuard, an OAuth 2.0 and OpenID Connect vulnerability
scanner and protector for RPs using Google OAuth 2.0 and OpenID Connect. It can
be used to protect user security and privacy even if RPs have not implemented
OAuth 2.0 or OpenID Connect correctly. We used OAuthGuard to check the security
and privacy properties of the 1,000 top-ranked websites supporting Google
sign-in; in particular OAuthGuard checked for five OAuth 2.0 or OpenID Connect
vulnerabilities. Of the 137 sites (from the 1000) that employ Google Sign-in,
69 were found to suffer from at least one serious vulnerability in their
implementation of OAuth 2.0 or OpenID Connect. OAuthGuard is able to protect
user security and privacy for 56 of these 69 vulnerable RPs, and provide a
warning to users of the other 13.


\bibliographystyle{plain}

\section*{The Appendix}
\label{appendix}
\begin{lstlisting}[caption={The OAuth 2.0 Request and Response metadata}, label={listing:oauth2metadata}]
// An OAuth 2.0 request metadata
// the requestURL and state in the request are trimmed for readability
IdP: "https://accounts.google.com"
IdPProtocol: "https:"
RP: "www.dropbox.com"
RPDomain: "dropbox.com"
RPProtocol: "https:"
clientID: "801668726815.apps.googleusercontent.com"
origin: null
redirectURI: "https://www.dropbox.com/google/authcallback"
referer: "https://www.dropbox.com/"
requestURL: "https://accounts.google.com/o/oauth2/auth"
responseType: "code"
scope: "https://www.google.com/m8/feeds email profile"
state: "ABAm_Lg53XmdhkeMTOmFKH5RULv2egJHsRXl9KHhp6Tazub"

// an OAuth 2.0 response metadata
// the referer, responseURL and state in the response are trimmed for readability
IdP: "google.com"
RPDomain: "dropbox.com"
RPHost: "www.dropbox.com"
RPProtocol: "https:"
access_token: ""
code: "4/gKfVUfaN5n-9tmo3RYnYActwrYWIXAwnsXRA7fcUl6E"
cookie: ""
data: ""
id_token: ""
method: "GET"
referer: "https://accounts.google.com/signin/oauth/oauthchooseaccount?"
responseURL: "https://www.dropbox.com/google/authcallback?"
state: "ABAm_Lg53XmdhkeMTOmFKH5RULv2egJHsRXl9KHhp6Tazub"
\end{lstlisting}

\end{document}